\documentclass[12pt]{article}
 \usepackage{epsfig}
 \def\be{\begin{equation}}
 \def\ee{\end{equation}}
 \def\bea{\begin{eqnarray}}
 \def\eea{\end{eqnarray}}
 \usepackage{graphicx}

 \catcode`\@=11
 \def\lsim{\mathrel{\mathpalette\@versim<}}
 \def\gsim{\mathrel{\mathpalette\@versim>}}
 \def\@versim#1#2{\vcenter{\offinterlineskip
 \ialign{$\m@th#1\hfil##\hfil$\crcr#2\crcr\sim\crcr } }}
 \catcode`\@=12

 \catcode`@=12
\begin{document}
 \thispagestyle{empty}
 \begin{flushright}
 UCRHEP-T574\\
 January 2017\
 \end{flushright}
 \vspace{0.6in}
 \begin{center}
 {\LARGE \bf Quartified Leptonic Color, Bound States,\\ 
 and Future Electron-Positron Collider\\}
 \vspace{1.2in}
 {\bf Corey Kownacki, Ernest Ma, Nicholas Pollard, Oleg Popov, and 
Mohammadreza Zakeri\\}
 \vspace{0.2in}
 {\sl Physics and Astronomy Department,\\ 
 University of California, Riverside, California 92521, USA\\}
 \end{center}
 \vspace{1.2in}

\begin{abstract}\
The $[SU(3)]^4$ quartification model of Babu, Ma, and Willenbrock (BMW), 
proposed in 2003, predicts a confining leptonic color $SU(2)$ gauge symmetry, 
which becomes strong at the keV scale.  It also predicts the existence of 
three families of half-charged leptons (hemions) below the TeV scale. 
These hemions are confined to form bound states which are not so easy to  
discover at the Large Hadron Collider (LHC).  However, just as $J/\psi$ 
and $\Upsilon$ appeared as sharp resonances in $e^-e^+$ colliders of the 
20th centrury, the corresponding 'hemionium' states are expected at a 
future $e^-e^+$ collider of the 21st century. 
\end{abstract}

 \newpage
 \baselineskip 24pt
\noindent \underline{\it Introduction}~:~
Fundamental matter consists of quarks and leptons, but why are they so 
different?  Both interact through the $SU(2)_L \times U(1)_Y$ electroweak 
gauge bosons $W^\pm, Z^0$ and the photon $A$, but only quarks interact 
through the strong force as mediated by the gluons of the unbroken (and 
confining) color $SU(3)$ gauge symmetry, called quantum chromodynamics (QCD). 
Suppose this is only true of the effective low-energy theory.  At high 
energy, there may in fact be three 'colors' of leptons transforming as 
a triplet under a leptonic color $SU(3)$ gauge symmetry.  Unlike QCD, 
only its $SU(2)_l$ subgroup remains exact, thus confining only two of the 
three 'colored' leptons, called 'hemions' in Ref.~\cite{bmw04} because they 
have $\pm 1/2$ electric charges, leaving the third ones free as the known 
leptons.

The notion of leptonic color was already discussed many years 
ago~\cite{fl90,flv91}, and its incorporation into $[SU(3)]^4$ appeared in 
Ref.~\cite{jv92}, but without full unification.  Its relevance today is 
threefold. (1) The $[SU(3)]^4$ quartification model~\cite{bmw04} of Babu, 
Ma, and Willenbrock (BMW) is non-supersymmetric, and yet achieves 
gauge-coupling unification at $4 \times 10^{11}$ GeV without endangering 
proton decay.  This unification of gauge couplings is only possible if 
the three families of hemions have masses below the TeV scale. Given 
the absence of experimental evidence for supersymmetry at the Large Hadron 
Collider (LHC) to date, this alternative scenario deserves a closer look. 
(2) The quartification scale determines the common gauge coupling for 
the $SU(2)_l$ symmetry.  Its extrapolation to low energy predicts that 
it becomes strong at the keV scale, in analogy to that of QCD becoming 
strong at somewhat below the GeV scale.  This may alter the thermal history 
of the Universe and allows the formation of 
gauge-boson bound states, the lightest of which is a potential 
warm dark-matter candidate~\cite{sz16}.  (3)  The hemions (called 
'liptons' previously~\cite{flv91}) have $\pm 1/2$ electric charges and are 
confined to form bound states by the $SU(2)_l$ 'stickons' in analogy to 
quarks forming hadrons through the $SU(3)_C$ gluons.  They have been 
considered previously~\cite{flv90} as technifermions responsible for 
electroweak symmetry breaking.  Their electroweak production at the LHC is 
possible~\cite{cfv12} but the background is large.  However, in a future 
$e^-e^+$ collider (ILC, CEPC, FCC-ee), neutral vector resonances of their 
bound states (hemionia) would easily appear, in analogy to the observations 
of quarkonia ($J/\psi$, $\Upsilon$) at past $e^-e^+$ colliders. 

\noindent \underline{\it The BMW model}~:~
Under the $[SU(3)]^4$ quartification gauge symmetry, quarks and leptons 
transform as $(3,\bar{3})$ in a moose chain linking $SU(3)_q$ to $SU(3)_L$ to 
$SU(3)_l$ to $SU(3)_R$ back to $SU(3)_q$ as depicted in Fig.~1.  
\begin{figure}[htb]
\vspace*{-3cm}
\hspace*{-3cm}
\includegraphics[scale=1.0]{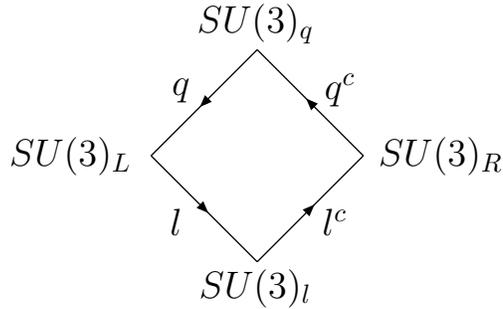}
\vspace*{-21.5cm}
\caption{Moose diagram of $[SU(3)]^4$ quartification.}
\end{figure}

\noindent Specifically,
\begin{eqnarray}
&& q \sim (3,\bar{3},1,1) \sim \pmatrix{d & u & h \cr d & u & h \cr d & u & h},
~~~ l \sim (1,3,\bar{3},1) \sim \pmatrix{x_1 & x_2 & \nu \cr y_1 & y_2 & e 
\cr z_1 & z_2 & N}, \\ 
&& l^c \sim (1,1,3,\bar{3}) \sim \pmatrix{x^c_1 & y_1^c & z_1^c \cr x_2^c & 
y_2^c & z_2^c \cr \nu^c & e^c & N^c}, ~~~  
q^c \sim (\bar{3},1,1,3) \sim \pmatrix{d^c & d^c & d^c \cr u^c & u^c & u^c 
\cr h^c & h^c & h^c}.
\end{eqnarray}
Below the TeV energy scale, the gauge symmetry is reduced~\cite{bmw04} to 
$SU(3)_C \times SU(2)_l \times SU(2)_L \times U(1)_Y$ with the particle 
content given in Table 1.
\begin{table}[htb]
\caption{Particle content of proposed model.}
\begin{center}
\begin{tabular}{|c|c|c|c|c|}
\hline
particles & $SU(3)_C$ & $SU(2)_l$ & $SU(2)_L$ & $U(1)_Y$ \\
\hline
$(u,d)_L$ & 3 & 1 & 2 & 1/6 \\
$u_R$ & $3$ & 1 & 1 & 2/3 \\
$d_R$ & $3$ & 1 & 1 & $-1/3$ \\
\hline
$(x,y)_L$ & 1 & 2 & 2 & 0 \\ 
$x_R$ & 1 & 2 & 1 & 1/2 \\ 
$y_R$ & 1 & 2 & 1 & $-1/2$ \\ 
\hline
$(\nu,l)_L$ & 1 & 1 & 2 & $-1/2$ \\
$\nu_R$ & 1 & 1 & 1 & 0 \\ 
$l_R$ & 1 & 1 & 1 & $-1$ \\
\hline
$(\phi^+,\phi^0)$ & 1 & 1 & 2 & $1/2$ \\
\hline
\end{tabular}
\end{center}
\end{table}
The electric charge $Q$ is given by $Q = I_{3L} + Y$ as usual.  The exotic 
$SU(2)_l$ doublets $x,y$ have $\pm 1/2$ charges, hence the name 
hemions.  Whereas the quarks and charged leptons must obtain masses 
through electroweak symmetry breaking, the hemions have invariant mass 
terms, i.e. $x_{1L} y_{2L} - x_{2L} y_{1L}$ and $x_{1R} y_{2R} - x_{2R} y_{1R}$.  
This is important because they are then allowed to be heavy without 
disturbing the electroweak oblique parameters $S,T,U$ which are highly 
constrained experimentally.  In the following, the mass terms from 
electroweak symmetry breaking, i.e. $\bar{x}_L x_R \bar{\phi}^0$ and 
$\bar{y}_L y_R \phi^0$, will be assumed negligible.

\noindent \underline{\it Gauge coupling unification and the leptonic color 
confinement scale}~:~
The renormalization-group evolution of the gauge couplings is dictated at 
leading order by
\begin{equation}
{1 \over \alpha_i(\mu)} - {1 \over \alpha_i(\mu')} = {b_i \over 2 \pi} 
\ln \left( {\mu' \over \mu} \right),
\end{equation}
where $b_i$ are the one-loop beta-function coefficients,
\begin{eqnarray}
b_C &=& -11 + {4 \over 3} N_F, \\ 
b_l &=& -{22 \over 3} + {4 \over 3} N_F, \\ 
b_L &=& -{22 \over 3} + 2 N_F + {1 \over 6} N_\Phi, \\ 
b_Y &=& {13 \over 9} N_F + {1 \over 12} N_\Phi.
\end{eqnarray}
The number of families $N_F$ is set to three, and the number of Higgs 
doublets $N_\Phi$ is set to two, as in the original BMW model.  Here we make a 
small adjustment by separating the three hemion families into two 
light ones at the electroweak scale $M_Z$ and one at a somewhat higher scale 
$M_X$.  We then input the values~\cite{pdg16}
\begin{eqnarray}
\alpha_C (M_Z) &=& 0.1185, \\ 
\alpha_L (M_Z) &=& (\sqrt{2}/\pi)G_F M_W^2 = 0.0339, \\ 
\alpha_Y (M_Z) &=& 2 \alpha_L (M_Z) \tan^2 \theta_W = 0.0204,
\end{eqnarray}
where $\alpha_Y$ has been normalized by a factor of 2 (and $b_Y$ by a factor 
of 1/2) to conform to $[SU(3)]^4$ quartification.  We find 
\begin{equation}
M_U = 4 \times 10^{11}~{\rm GeV}, ~~~ \alpha_U = 0.0301, ~~~ M_X = 
486~{\rm GeV}.
\end{equation}
We then use $b_l$ to extrapolate back to $M_Z$ and obtain $\alpha_l(M_Z) 
= 0.0469$.  Below the electroweak scale, the evolution of $\alpha_l$ 
comes only from the stickons and it becomes strong at about 1 keV. 
Hence 'stickballs' are expected at this confinement mass scale.  Unlike 
QCD where glueballs are heavier than the $\pi$ mesons so that 
they decay quickly, the stickballs are so light that they could decay 
only to lighter stickballs or to photon pairs through their interactions 
with hemions.

\noindent \underline{\it Thermal history of stickons}~:~
At temperatures above the electroweak symmetry scale, the hemions are 
active and the stickons ($\zeta$) are in thermal equilibrium with the 
standard-model particles.  Below the hemion mass scale, the stickon 
interacts with photons through $\zeta \zeta \to \gamma \gamma$ 
scattering with a cross section
\begin{equation}
\sigma \sim {9 \alpha^2 \alpha_l^2 T^6 \over 16 M^8_{eff}}.
\end{equation}
The decoupling temperature of $\zeta$ is then obtained by matching the 
Hubble expansion rate
\begin{equation}
H = \sqrt {(8 \pi/3) G_N (\pi^2/30) g_* T^4}
\end{equation}
to $[6\zeta(3)/\pi^2] T^3 \langle \sigma v \rangle$.  Hence 
\begin{equation}
T^{14} \sim {2^{8} \over 3^8} \left( {\pi^7 \over 5 [\zeta(3)]^2} \right) 
{G_N g_* M^{16}_{eff} \over \alpha^4 \alpha_l^4},
\end{equation}
where $6 M_{eff}^{-4} = \sum (M^i_{xy})^{-4}$.  For $M_{eff} = 110$ GeV 
and $g_* = 92.25$ which includes all particles with masses up to a few GeV,  
$T \sim 6.66$ GeV.  Hence the contribution of 
stickons to the effective number of neutrinos at the time of big bang 
nucleosynthesis (BBN) is given by~\cite{jt13}
\begin{equation}
\Delta N_\nu = {8 \over 7} (3) \left( {10.75 \over 92.25} \right)^{4/3} 
= 0.195,
\end{equation}
compared to the value $0.50 \pm 0.23$ from a recent analysis~\cite{ns15}.  
The most recent PLANCK measurement~\cite{planck16} coming from the cosmic 
microwave background (CMB) is 
\begin{equation}
N_{eff} = 3.15 \pm 0.23.
\end{equation}
However, at the time of photon decoupling, the stickons have disappeared, 
hence $N_{eff} = 3.046$ as in the SM.  This is discussed in more detail 
below.

\noindent \underline{\it Formation and decay of stickballs}~:~
As the Universe further cools below a few keV, leptonic color goes through 
a phase transition and stickballs are formed.  If the lightest 
stickball $\omega$ is stable, it may be a candidate for warm dark matter.  
It has strong self-interactions and the $3 \to 2$ process determines 
its relic abundance.  Following Ref.~\cite{cmh92} and using Ref.~\cite{sz16}, 
we estimate that it is overproduced by a factor of about 3.  However, 
$\omega$ is not absolutely stable.  It is allowed to mix with a scalar 
bound state of two hemions which would decay to two photons.  We assume 
this mixing to be $f_\omega m_\omega/M_{xy}$, so that its decay rate is 
given by
\begin{equation}
\Gamma (\omega \to \gamma \gamma) = {9 \alpha^2 f_\omega^2 m_\omega^5 \over 
64 \pi^3 M_{eff}^4},
\end{equation}
where $M_{eff}$ is now defined by $6 M^{-2}_{eff} = \sum (M^i_{xy})^{-2}$. 
Setting $m_\omega = 5$ keV to be above the astrophysical bound of 4 keV from 
Lyman $\alpha$ forest observations~\cite{bpymv16} and $M_{eff} = 150$ GeV,
its lifetime is estimated to be $4.4 \times 10^{17}s$ for $f_\omega=1$.  
This is exactly the age of the Universe, and it appears that $\omega$ 
may be a candidate for dark matter after all.  However, CMB measurements 
constrain~\cite{sw16} a would-be dark-matter lifetime to be greater than 
about $10^{25}s$, and $x$-ray line measurements in this mass range 
constrain~\cite{mpq16} it to be greater than $10^{27}s$, so this scenario 
is ruled out.  On the other hand, if 
$m_\omega = 10$ keV, then the $\omega$ lifetime is $1.4 \times 10^{16}s$, 
which translates to a fraction of $2 \times 10^{-14}$ of the initial 
abundance of $\omega$ to remain at the present Universe.  Compared to the 
upper bound of $10^{-10}$ for a lifetime of $10^{16}s$ given in 
Ref.~\cite{sw16}, this is easily 
satisfied, even though $\omega$ is overproduced at the leptonic color 
phase transition by a factor of 3.
 
At the time of photon decoupling, the $SU(2)_l$ sector contributes 
no additional relativistic degrees of freedom, hence $N_{eff}$ remains 
the same as in the SM, i.e. 3.046, coming only from neutrinos. 
In this scenario, $\omega$ is not dark matter.  However, there are many 
neutral scalars and fermions in the BMW model which are not being 
considered here.  They are naturally very heavy, but some may be light 
enough and stable, and be suitable as dark matter.

\noindent \underline{\it Revelation of leptonic color at future $e^-e^+$ 
colliders}~:~
Unlike quarks, all hemions are heavy.  Hence the lightest bound state is 
likely to be at least 200 GeV.  Its cross section through electroweak 
production at the LHC is probably too small for it to be discovered. 
On the other hand, in analogy to the observations of $J/\psi$ and $\Upsilon$ 
at $e^-e^+$ colliders of the last century, the resonance production of 
the corresponding neutral vector bound states (hemionia) of these hemions 
is expected at a future $e^-e^+$ collider (ILC, CEPC, FCC-ee) with sufficient 
reach in total center-of-mass energy.  Their decays will be distinguishable 
from heavy quarkonia (such as toponia) experimentally.

The formation of hemion bound states is analogous to that of 
QCD.  Instead of one-gluon exchange, the Coulomb potential binding a 
hemion-antihemion pair comes from one-stickon exchange.  The difference is 
just the change in an SU(3) color factor of 4/3 to an SU(2) color factor 
of 3/4.  The Bohr radius is then $a_0 = [(3/8) \bar{\alpha}_l m]^{-1}$, 
and the effective $\bar{\alpha}_l$ is defined by
\begin{equation} 
\bar {\alpha}_l = \alpha_l (a_0^{-1}).
\end{equation}
Using Eqs.~(3) and (5), and $\alpha_l (M_Z) = 0.047$ with $m=100$ GeV, 
we obtain $\bar{\alpha}_l = 0.059$ and $a_0^{-1} = 2.2$ GeV.  Consider 
the lowest-energy vector bound state $\Omega$ of the lightest hemion 
of mass $m=100$ GeV.  In analogy to the hydrogen atom, its binding energy 
is given by
\begin{equation}
E_b = {1 \over 4} \left( {3 \over 4} \right)^2 \bar{\alpha}_l^2 m 
= 0.049~{\rm GeV},
\end{equation}
and its wavefunction at the origin is
\begin{equation}
|\psi(0)|^2 = {1 \over \pi a_0^3} = 3.4~{\rm GeV}^3.
\end{equation}
Since $\Omega$ will appear as a narrow resonance at a future $e^-e^+$ 
collider, its observation depends on the integrated cross section over 
the energy range $\sqrt{s}$ around $m_\Omega$:
\begin{equation}
\int d \sqrt{s} ~\sigma (e^- e^+ \to \Omega \to X) = {6 \pi^2 \over 
m_\Omega^2} {\Gamma_{ee} \Gamma_X \over \Gamma_{tot}},
\end{equation}
where $\Gamma_{tot}$ is the total decay width of $\Omega$, and $\Gamma_{ee}$, 
$\Gamma_X$ are the respective partial widths.

Since $\Omega$ is a vector meson, it couples to both the photon and $Z$ boson 
through its constituent hemions.  Hence it will decay to $W^- W^+$, 
$q \bar{q}$, $l^- l^+$, and $\nu \bar{\nu}$.  Using
\begin{equation}
\langle 0 | \bar{x} \gamma^\mu x | \Omega \rangle = \epsilon_\Omega^\mu 
\sqrt{8 m_\Omega} |\psi(0)|,
\end{equation}
the $\Omega \to e^- e^+$ decay rate is given by
\begin{equation}
\Gamma (\Omega \to \gamma,Z \to e^- e^+) = {2 m_\Omega^2 \over 3 \pi} 
(|C_V|^2 + |C_A|^2) |\psi(0)|^2,
\end{equation}
where
\begin{eqnarray}
C_V &=& {e^2 (1/2) (-1) \over m_\Omega^2} + {g_Z^2 (-\sin^2 \theta_W/4)  
[(-1+4\sin^2 \theta_W)/4] \over m_\Omega^2 - M_Z^2}, \\ 
C_A &=& {g_Z^2 (-\sin^2 \theta_W/4)  (1/4) \over m_\Omega^2 - M_Z^2}.
\end{eqnarray}
In the above, $\Omega$ is assumed to be composed of the singlet hemions 
$x_R$ and $y_R$ with invariant mass term $x_{1R} y_{2R} - x_{2R} y_{1R}$ 
(case A).  Hence $\Gamma_{ee} = 43$ eV.  If $\Omega$ comes instead from 
$x_L$ and $y_L$ with invariant mass term $x_{1L} y_{2L} - x_{2L} y_{1L}$ 
(case B), then the factor $(-\sin^2 \theta_W/4)$ in $C_V$ and $C_A$ is 
replaced with $(\cos^2 \theta_W/4)$ and $\Gamma_{ee} = 69$ eV.  Similar 
expressions hold for the other fermions of the Standard Model (SM).

For $\Omega \to W^- W^+$, the triple $\gamma W^- W^+$ and $Z W^- W^+$ vertices 
have the same structure.  The decay rate is calculated to be
\begin{equation}
\Gamma (\Omega \to \gamma,Z \to W^- W^+) = {m_\Omega^2 (1-r)^{3/2} \over 
6 \pi r^2} \left( 4 + 20r + 3 r^2 \right) 
C_W^2 |\psi(0)|^2,
\end{equation}
where $r = 4M_W^2/m_\Omega^2$ and 
\begin{equation}
C_W = {e^2 (1/2) \over m_\Omega^2} + {g_Z^2 (-\sin^2 \theta_W/4) \over 
m_\Omega^2 - M_Z^2}
\end{equation}
in case A.  Because of the accidental cancellation of the two terms in 
the above, $C_W$ turns out to be very small.  Hence $\Gamma_{WW} = 3.2$ eV.  
In addition to the $s-$channel decay of $\Omega$ to $W^-W^+$ through 
$\gamma$ and $Z$, there is also a $t-$channel electroweak contribution 
in case B because $x_L$ and $y_L$ form an electroweak doublet.  
Replacing $(-\sin^2 \theta_W/4)$ with $(\cos^2 \theta_W/4)$ in $C_W$, 
and adding this contribution, we obtain
\begin{eqnarray}
\Gamma (\Omega \to W^-W^+) &=& {m_\Omega^2 (1-r)^{3/2} \over 6 \pi r^2} 
[(4 + 20r + 3r^2) C_W^2 \nonumber \\
&+& 2r(10 + 3r) C_W D_W + r(8 - r) D_W^2] |\psi(0)|^2,
\end{eqnarray}
where
\begin{equation}
D_W = {-g^2 \over 4 (m_\Omega^2 - 2 M_W^2)}.
\end{equation}
Thus a much larger $\Gamma_{WW} = 190$ eV is obtained. 
For $\Omega \to ZZ$, there is only the $t-$channel contribution, i.e.
\begin{equation}
\Gamma(\Omega \to ZZ) = {m_\Omega^2 (1-r_Z)^{5/2} \over 3 \pi r_Z} D_Z^2 
|\psi(0)|^2,
\end{equation}
where $r_Z = 4M_Z^2/m_\Omega^2$ and $D_Z = g_Z^2 \sin^4 \theta_W /
4(m_\Omega^2-2m_Z^2)$ in case A, with $\sin^4 \theta_W$ replaced by 
$\cos^4 \theta_W$ in case B.  Hence $\Gamma_{ZZ}$ is negligible in case A 
and only 2.5 eV in case B.

The $\Omega$ decay to two stickons is forbidden by charge conjugation. 
Its decay to three stickons is analogous to that of quarkonium to three 
gluons.  Whereas the latter forms a singlet which is symmetric in $SU(3)_C$, 
the former forms a singlet which is antisymmetric in $SU(2)_l$.  However, 
the two amplitudes are identical because the latter is symmetrized with 
respect to the exchange of the three gluons and the former is antisymmetrized 
with respect to the exchange of the three stickons.  Taking into account the 
different color factors of $SU(2)_l$ versus $SU(3)_C$, the decay rate of 
$\Omega$ to three stickons and to two stickons plus a photon are given by
\begin{eqnarray}
\Gamma (\Omega \to \zeta \zeta \zeta) &=& {16 \over 27} (\pi^2-9) {\alpha_l^3 
\over m_\Omega^2} |\psi(0)|^2, \\ 
\Gamma (\Omega \to \gamma \zeta \zeta) &=& {8 \over 9} (\pi^2-9) {\alpha 
\alpha_l^2 \over m_\Omega^2} |\psi(0)|^2. 
\end{eqnarray}
Hence $\Gamma_{\zeta \zeta \zeta} = 4.5$ eV and $\Gamma_{\gamma \zeta \zeta} 
= 1.1$ eV.
The integrated cross section of Eq.~(21) for $X = \mu^- \mu^+$ is then 
$3.8 \times 10^{-33}$ cm$^2$-keV in case A and $2.1 \times 10^{-33}$ 
cm$^2$-keV 
in case B.  For comparison, this number is $7.9 \times 10^{-30}$ cm$^2$-keV 
for the $\Upsilon(1S)$.  At a high-luminosity $e^- e^+$ collider, it should 
be feasible to make this observation.  Table 2 summarizes all the partial 
decay widths.

\begin{table}[htb]
\caption{Partial decay widths of the hemionium $\Omega$.}
\begin{center}
\begin{tabular}{|c|c|c|}
\hline
Channel & Width (A) & Width (B) \\
\hline
$\nu \bar{\nu}$ & 11 eV & 123 eV \\
\hline
$e^- e^+$ & 43 eV & 69 eV \\
$\mu^- \mu^+$ & 43 eV & 69 eV \\
$\tau^- \tau^+$ & 43 eV & 69 eV \\
\hline
$u \bar{u}$ & 50 eV & 175 eV \\
$c \bar{c}$ & 50 eV & 175 eV \\
\hline
$d \bar{d}$ & 10 eV & 147 eV \\
$s \bar{s}$ & 10 eV & 147 eV \\
$b \bar{b}$ & 10 eV & 147 eV \\
\hline
$W^- W^+$ & 3.2 eV & 190 eV \\
$Z Z$ & 0.02 eV & 2.5 eV \\
\hline
$\zeta \zeta \zeta$ & 4.5 eV & 4.5 eV \\ 
$\zeta \zeta \gamma$ & 1.1 eV & 1.1 eV \\
\hline
sum & 279 eV & 1319 eV \\
\hline
\end{tabular}
\end{center}
\end{table}

\noindent \underline{\it Discussion and outlook}~:~
There are important differences between QCD and QHD (quantum hemiodynamics). 
In the former, because of the existence of light $u$ and $d$ quarks, it is 
easy to pop up $u \bar{u}$ and $d \bar{d}$ pairs from the QCD vacuum.  
Hence the production of open charm in an $e^- e^+$ collider is described 
well by the fundamental process $e^- e^+ \to c \bar{c}$.  In the latter, 
there are no light hemions.  Instead it is easy to pop up the light 
stickballs from the QHD vacuum.  As a result, just above the threshold 
of making the $\Omega$ resonance, the many-body production of $\Omega$ + 
stickballs becomes possible.  This cross section is presumably also well 
described by the fundamental process $e^- e^+ \to x \bar{x}$.  In case A, 
the cross section is given by
\begin{eqnarray}
\sigma (e^- e^+ \to x \bar{x}) &=& {2 \pi \alpha^2 \over 3} 
\sqrt{1-{4m^2 \over s}}\left[ 
{(s + 2m^2) \over s^2} + {x_W^2 \over 2 (1-x_W)^2} {(s-m^2) \over (s-m_Z^2)^2} 
\right. 
\nonumber \\ &+& \left. {x_W \over (1-x_W)} {(s-m^2) \over s(s-m_Z^2)} - 
{(1-4x_W) \over 4(1-x_W)} {m^2 \over s(s-m_Z^2)} \right],
\end{eqnarray}
where $x_W = \sin^2 \theta_W$ and $s = 4E^2$ is the square of the 
center-of-mass energy.  In case B, it is
\begin{eqnarray}
\sigma (e^- e^+ \to x \bar{x}) &=& 
{2 \pi \alpha^2 \over 3} \sqrt{1-{4m^2 \over s}} \left[ 
{(s + 2m^2) \over s^2} + {(s-m^2) \over 2(s-m_Z^2)^2} \right. 
\nonumber \\ &-& \left. {(s-m^2) \over s(s-m_Z^2)} + 
{(1-4x_W) \over 4 x_W} {m^2 \over s(s-m_Z^2)} \right].
\end{eqnarray}
Using $m=100$ GeV and $s = (250~{\rm GeV})^2$ as an example, we find 
these cross sections to be 0.79 and 0.44 pb respectively.

In QCD, there are $q \bar{q}$ bound states which are bosons, and $qqq$ 
bound states which are fermions.  In QHD, there are only bound-state 
bosons, because the confining symmetry is $SU(2)_l$.  Also, unlike 
baryon (or quark) number in QCD, there is no such thing as hemion number 
in QHD, because $y$ is effectively $\bar{x}$.  This explains why there are 
no stable analog fermion in QHD such as the proton in QCD.

The SM Higgs boson $h$ couples to the hemions, but these Yukawa couplings 
could be small, because hemions have invariant masses themselves as 
already explained.  So far we have assumed these couplings to be negligible. 
If not, then $h$ may decay to two photons and two stickons through a loop 
of hemions.  This may show up in precision Higgs studies as a deviation 
of $h \to \gamma \gamma$ from the SM prediction.  It will also imply a 
partial invisible width of $h$ proportional to this deviation.  Neither 
would be large effects and that is perfectly consistent with present data.

The absence of observations of new physics at the LHC is a possible 
indication that fundamental new physics may not be accessible using 
the strong interaction, i.e. quarks and gluons.  It is then natural to 
think about future $e^- e^+$ colliders.  But is there some fundamental 
issue of theoretical physics which may only reveal itself there? and not 
at hadron colliders?  The BMW model is one possible answer.  It assumes 
a quartification symmetry based on $[SU(3)]^4$.  It has gauge-coupling 
unification without supersymmetry, but requires the existence of new 
half-charged fermions (hemions) under a confining $SU(2)_l$ leptonic 
color symmetry, with masses below the TeV scale.  It also predicts 
the $SU(2)_l$ confining scale to be keV, so that stickball bound states 
of the vector gauge stickons are formed.  These new particles have no 
QCD interactions, but hemions have electroweak couplings, so they 
are accessible in a future $e^- e^+$ collider, as described in this 
paper.

\noindent \underline{\it Acknowledgement}~:~
This work was supported in part by the U.~S.~Department of Energy Grant 
No. DE-SC0008541.

\newpage
\bibliographystyle{unsrt}

\end{document}